\newcommand{\be}{\begin{equation}}
\newcommand{\ee}{\end{equation}}

\newcommand{\diag}{{\rm diag}}

\newcommand{\al}{\alpha}

\newcommand{\p}{\partial}

\newcommand{\hook}{{\setlength{\unitlength}{10pt}   
                   \begin{picture}(.833,.8)
                   \put(.15,.08){\line(1,0){.35}}
                   \put(.5,.08){\line(0,1){.5}}
                   \end{picture}}}

\newtheorem{definition}{Definition}
\newtheorem{theorem}{Theorem}

\newtheorem{proposition}[theorem]{Proposition}
\newtheorem{example}{Example}

\documentclass{article}
\usepackage{amsfonts}
\usepackage{amssymb}
\usepackage{amsbsy}

\begin{document}
\begin{flushright}
\end{flushright}
\vspace{0.2cm}
\begin{center}
\begin{Large}
\textbf{The twistor theory of Whitham hierarchy}
\end{Large}\\
\vspace{1.0cm}

M. Y. Mo$^{\sharp}$ \footnote{e-mail: mo@crm.umontreal.ca}
\bigskip\\
$^\sharp$ Centre de recherche math\'ematiques, Universit\'e\ de
Montr\'eal.
\bigskip\\
{\bf Abstract}
\end{center}
We have generalized the approach in of Dunajski, Mason and Tod
\cite{dmt} and established a 1-1 correspondence between a solution
of the universal Whitham hierarchy \cite{k94} and a twistor space.
The twistor space consists of a complex surface and a family of
complex curves together with a meromorphic 2-form. The solution of
the Whitham hierarchy is given by deforming the curve in the
surface. By treating the family of algebraic curves in
$\mathbb{CP}^1\times\mathbb{CP}^1$ as a twistor space, we were
able to express the deformations of the isomonodromic spectral
curve in terms of the deformations generated by the Whitham
hierarchy.

\section{Introduction}\label{se:introduction}
The Whitham hierarchy originates from the study of the
'dispersionless limit' of integrable systems. The typical setup
involves an introduction of a small parameter $\epsilon$ and a
suitable rescaling of the time variables. For example, if one
rescale the times $x$, $t$ to $\tilde{x}=\epsilon x$,
$\tilde{t}=\epsilon t$ in the KdV equation
\begin{eqnarray*}
\p_tu=6u\p_xu-\p_x^3u
\end{eqnarray*}
one obtains the KdV equation with small dispersion
\begin{eqnarray}\label{eq:sdisp}
\tilde{\p}_tu(\tilde{x},\tilde{t},\epsilon)=6u\tilde{\p}_xu(\tilde{x},\tilde{t},\epsilon)-
\epsilon^2\tilde{\p}_x^3u(\tilde{x},\tilde{t},\epsilon)
\end{eqnarray}
where $\tilde{\p}_x$ and $\tilde{\p}_t$ denote the derivatives
with respect to the tilded variables.

The problem is to study the behavior of the solution
$u(\tilde{x},\tilde{t},\epsilon)$ as $\epsilon\rightarrow 0$. In
general, the solution contains parts that oscillates rapidly as
$\epsilon\rightarrow 0$ and the limit does not exist.

However, the weak limit of $u(\tilde{x},\tilde{t},\epsilon)$ can
be studied. This is like `averaging out' the oscillatory part and
study the limit, and such an averaging process is called the
Whitham averaging. \cite{dn83}, \cite{dn89}, \cite{gp73},
\cite{fm76}, \cite{ffm80}, \cite{k88}, \cite{k91}, \cite{ll83a},
\cite{ll83b}, \cite{ll83c}, \cite{wh}

There is an interesting connection between the small dispersion
KdV and deformations of Riemann surfaces. \cite{ffm80}

A finite gap solution of the KdV equation can be expressed in
terms of theta functions of a genus $g$ algebraic curve (the
spectral curve). The spectral curve is fixed under the evolution
of the finite gap solution. \cite{dmn76}, \cite{k77a}, \cite{k77b}
In many applications, a solution $u(x,t)$ of the KdV equation can
be approximated by a finite gap solution locally. However, the
approximation becomes bad as $(x,t)$ grows large and the spectral
curve that classifies the finite gap solution varies slowly with
respect to $(x,t)$. To study the variations of the spectral curve,
one could think of the spectral curve as depending on the 'slow
times' $(\tilde{x}=\epsilon x, \tilde{t}=\epsilon t)$ and study
the dependence of $u$ on $(\tilde{x},\tilde{t})$ in
(\ref{eq:sdisp}) as $\epsilon\rightarrow 0$.

In this case, The weak limit of $u(\tilde{x},\tilde{t},\epsilon)$
can be described by $2g+1$ functions that satisfy the $g$-phase
Whitham equations \cite{wh}
\begin{eqnarray}\label{eq:modul}
\tilde{\p}_tu_i-v_{g,i}(u_1,\ldots,u_{2g+1})\tilde{\p}_xu_i=0,\quad
1=1,2,\ldots, 2g+1
\end{eqnarray}
where $v_{g,i}(u_1,\ldots,u_{2g+1})$ depends on $u_i$ on complete
hyperelliptic integrals of genus $g$. These equations can in fact
be written in a more compact form
\begin{eqnarray}\label{eq:algmodul}
\tilde{\p}_tdp=\tilde{\p}_xdq
\end{eqnarray}
where $dp$ and $dq$ are certain meromorphic 1-forms that are
normalized on the spectral curve with respect to a choice of a
canonical basis of cycles. To be precise, the spectral curve in
this case is a hyperelliptic curve
\begin{eqnarray*}
w^2=R(z)=\prod_{i=1}^{2g+1}(z-r_i)
\end{eqnarray*}
and $dp$, $dq$ are 1-forms of the form
\begin{eqnarray*}
dp&=&{{\sum_{i=1}^gc_iz^i}\over{\sqrt{R(z)}}} \\
dq&=&{{\sum_{i=1}^{g+1}d_iz^i}\over{\sqrt{R(z)}}} \\
\int_{r_{2i}}^{r_{2i-1}}dp&=&\int_{r_{2i}}^{r_{2i-1}}dq=0
\end{eqnarray*}
To obtain (\ref{eq:modul}) from (\ref{eq:algmodul}), we expand
$dp$ and $dq$ in terms of a $z$ and coefficients would satisfy
(\ref{eq:modul}). \cite{fm76}, \cite{ffm80}, \cite{dn83},
\cite{k88}

Krichever has studied equations of the type (\ref{eq:algmodul}) on
more general algebraic curves that cover the dispersionless limit
of other integrable systems and introduced the notion of the
universal Whitham hierarchy \cite{k94}. The universal Whitham
hierarchies, like (\ref{eq:algmodul}), are expressed in terms of
meromorphic and holomorphic 1-forms on Riemann surfaces. It has
many interesting applications to 2-D topological quantum field
theory, Frobenius manifolds and string theory. \cite{dub92},
\cite{dub97}, \cite{dub98}, \cite{dub99}, \cite{k91}, \cite{k94},
\cite{wi91}

In the case where the curve in the question is of genus 0,
Dunajski, Mason and Tod has studied the dispersionless KP (dKP)
equation from a point of view of twistor theory, in which a
solutions of the dKP equation is described by a family of rational
curves in a complex surface and a meromorphic 2-form on the
surface. This reveals interesting relations between the dKP
equation and the Einstein-Weyl metric. \cite{dun}, \cite{dt},
\cite{dmt}. In \cite{gmm03}, special solutions of the dKP equation
was constructed by using twistor methods.

In this paper, we have generalized the construction of \cite{dmt}
to the case where the curves are of arbitrary genus. We have
established a one-one correspondence between a solution of the
universal Whitham hierarchy and a twistor space, which is a
complex surface with a family of genus $g$ curves in it. To be
precise, we have the following
\begin{definition}
A twistor space ${\mathcal{T}}$ of the truncated universal Whitham
hierarchy consists of:

1. A 2-dimensional complex manifold ${\mathcal{T}}$ with a
meromorphic 2-form $\Pi$ which is singular on a divisor $D$,

2. A family of genus-g embedded curves
$\left\{\Sigma_{g,t}\right\}$ and canonical basis of cycles on
each curve,

3. A covering $V_{\beta}$ of a neighborhood $U$ of the singular
divisor $D$, and local coordinates $k_{\beta}^{-1}$ on $V_{\beta}$
such that $k_{\beta}^{-1}$ has zeroes of order 1 at $D$. A fix
point $P_1$ on each curve $\Sigma_t$ such that $P_1\in D$.
\end{definition}
and
\begin{theorem}
There is a one-one correspondence between a solution of the
Whitham hierarchy and a twistor space ${\mathcal{T}}$ defined
above.
\end{theorem}
The idea is that as the curve moves around in the twistor space,
the values of the 2-form $\Pi$ evaluated on different curves would
give the values of the solution at different times.

We have also studied the following problem which relates the
isomonodromic spectral curve to the Whitham hierarchy. Consider
the isomonodromic deformations of the system of linear ODE
\begin{eqnarray*}
{{dY}\over{dz}}=A(z)Y
\end{eqnarray*}
where $A$ is a $n\times n$ matrix such that
\begin{eqnarray*}
A(z)=G^T(\Delta-z)^{-1}F+C
\end{eqnarray*}
where $G$, $F$ are $n\times r$ matrices constant in $z$,
$\Delta=\diag(\al_1,\ldots,\al_r)$ and $C=\diag(c_1,\ldots,c_n)$.

Define the \it dual isomonodromic spectral curve \rm \cite{h94},
\cite{sw} of this isomonodromic system by
\begin{eqnarray*}
\det \left[ \mathbb{M}- \pmatrix{
  z & 0 \cr
  0 & w\cr} \right]=0
\end{eqnarray*}
These spectral curves are known to vary as the matrix $A(z)$
deforms isomonodromically.

The family of spectral curves is a family of algebraic curves in
$\mathbb{CP}^1\times\mathbb{CP}^1$ defined by polynomials $P(z,w)$
that has degree at most $n$ in $w$ and $r$ in $z$. Since the
family of curves defined by all polynomials $P(z,w)$ with degree
at most $n$ in $w$ and $r$ in $z$ forms a twistor space for the
Whitham hierarchy, we can compare the changes of the spectral
curve induced by isomonodromic deformations and the changes
induced by the Whitham hierarchy. This problem was posted by
Takesaki in \cite{ta98a}, \cite{ta98b} in a slightly different
form. We were able to derive formulas that express the
isomonodromic flows in terms of the Whitham flows in proposition
\ref{pro:isowhit}.

This paper is organized as follows. In section \ref{sec:Whitham}
we will give a brief introduction to the universal Whitham
hierarchy and established some notations. In section \ref{sec:
twistorwhit} we will construct the twistor space for the Whitham
hierarchy and in section \ref{se:exwhitham} we will give some
examples and study the isomonodromic spectral curves.

\section{The Universal Whitham hierarchy}\label{sec:Whitham}
In this section we will give a brief introduction to the universal
Whitham hierarchy in \cite{k94}.

Let $M_{g,N,n_{\al}}$ be the moduli space of genus $g$ curves
$\Gamma_{g,q}$ with $N$ punctures $P_{\al}\left(q\right)$, fixed
$n_{\al}$-jets of local coordinates $k_{\al}^{-1}\left(q\right)$
in neighborhoods of $P_{\al}\left(q\right)$ and canonical basis of
cycles $a_i\left(q\right)$, $b_i\left(q\right)$. The marked
points, local coordinates and basis change from curve to curve,
but the genus $g$ of the curves, the total number of marked points
$N$ and the numbers $n_{\al}$ do not vary between curves. Let
$\sum_{\al}\left(n_{\al}+1\right)=H$ then the dimension of the
moduli space $M_{g,N,n_{\al}}$ is $3g-2+H$.

That is, we have
\begin{eqnarray*}
M_{g,N}=\left\{ \Gamma_{g,q} ,P_{\al}(q), k_{\al}^{-1}(q), a_i(q),
b_i(q) \in H_1 \left(\Gamma_{g,q}, \mathbb{Z}\right) \right\}
\end{eqnarray*}
The tautological bundle $\hat{M}_{g,N,n_{\al}} \rightarrow
M_{g,N,n_{\al}}$ is the fibre bundle over $M_{g,N,n_{\al}}$ with
fibre at each point $q$ of $M_{g,N,n_{\al}}$ being the curve
$\Gamma_{g,q}$. The curves $\Gamma_{g,q}$ at different points are
not holomorphic to each other in general.

We consider each fibre $\Gamma_{g,q}$ as a genus $g$ Riemann
surface with canonical basis
$\left\{a_i\left(q\right),b_i\left(q\right)\right\}$ of cycles,
marked points $P_{\al}\left(q\right)$ and local coordinates
$k_{\al}^{-1}\left(q\right)$ near $P_{\al}\left(q\right)$. We now
define various meromorphic 1-forms
$d\Omega_{\mu,\nu}\left(q\right)$ on these Riemann surfaces
according to the following
\begin{definition}\label{de:whithamforms}
The meromorphic 1-forms $d\Omega_A$ where $A=\left(\mu,\nu\right)$
a double index are defined by

1. $d\Omega_{\al ,i}\left(q\right)$ are meromorphic 1-forms of the
second kind which are holomorphic outside $P_{\al}\left(q\right)$
and have the form

\begin{eqnarray}\label{eq:omegaai}
d\Omega_{\al ,i}\left(q\right)=d\left(k_{\al}^i\left(q\right)+
O\left(k_{\al}^{-1}\left(q\right)\right)\right)
\end{eqnarray}
near $P_{\al}\left(q\right)$, where $i=1, \ldots ,n_{\al}$.

2. $d\Omega_{\al ,0}\left(q\right)$, $\al \neq 1$ are meromorphic
1-forms of the third kind with residues 1 and -1 at
$P_1\left(q\right)$ and $P_{\al}\left(q\right)$ respectively, and
they look like
\begin{eqnarray}\label{eq:omegaa0}
d\Omega_{\al
,0}&=&dk_{\al}\left(q\right)\left(k_{\al}^{-1}\left(q\right)
+O\left(k_{\al}^{-1}\left(q\right)\right)\right)
\nonumber \\
&-&dk_1\left(q\right)\left(k_1^{-1}\left(q\right)
+O\left(k_1^{-1}\left(q\right)\right)\right)
\end{eqnarray}
near $P_{\al}\left(q\right)$ and $P_1\left(q\right)$.

3. $d\Omega_{h,k}\left(q\right)$, ($k=1, \ldots , g$ and $h$
fixed) are the normalized holomorphic 1-forms
\begin{eqnarray}\label{eq:omegahk}
\oint_{a_i\left(q\right)}
d\Omega_{h,k}\left(q\right)=\delta_{i,k}, \quad i,k=1, \ldots, g
\end{eqnarray}
4. The differentials in 1. and 2. are uniquely determined by the
normalization conditions
\begin{eqnarray}\label{eq:normalcond}
\oint _{a_i\left(q\right)} d\Omega_A\left(q\right) =0
\end{eqnarray}
where the index $A=\left(\al, i\right)$.
\end{definition}

We now fix an index $\left(\al, i\right)=\left(1,1\right)$ and
denote $d\Omega_{1,1}\left(q\right)$ by $dp\left(q\right)$. Let
the multi-valued function $p\left(q\right)$ be defined by the
abelian integral of $dp\left(q\right)$
\begin{eqnarray}\label{eq:coordp}
p\left(P,q\right)=\int^P dp\left(P,q\right)
\end{eqnarray}
where $P\in \Gamma_{g,q}$. We will now suppress the dependence of
$q$ and bear in mind that everything depends on the point $q$ of
the moduli space. We can expand the local coordinates $k_{\al}$ in
terms of the function $p$ near each marked point $P_{\al}$.
\begin{eqnarray}\label{eq:expandjet}
k_{\al}=\sum_{s=-1}^{\infty}\nu_{\al,s}\left(p-p_{\al}\right)^s,
\quad \al\neq 1
\end{eqnarray}
where $p_{\al}=p\left(P_{\al}\right)$, and for $\al=1$
\begin{eqnarray}
k_1=p+\sum_{s=1}^{\infty}\nu_{1,s}p^s
\end{eqnarray}
We can now use the parameters
\begin{eqnarray}\label{eq:modcoord}
\Big\{p_{\al}&=&p\left(P_{\al}\right),\nu_{\al,s},\al=1,\ldots,N,s=-1,\ldots,n_{\al}\Big\}
\nonumber \\
\sigma_s&=&p\left(\Sigma_s\right), \quad dp\left(\Sigma_s\right)=0, \quad s=1,\ldots,2g-2 \\
U_i^p&=&\oint_{b_i}dp, \quad i=1,\ldots, g\nonumber
\end{eqnarray}
as coordinates on the moduli space $M_{g,N,n_{\al}}$. Note that we
only need the coefficients $\nu_{\al,s}$ up to $s=n_{\al}$ as we
are considering $n_{\al}$-jets, which means that terms of order
$k_{\al}^i$, $i> n_{\al}$ are all equivalent and does not make any
difference.

Suppose we have some \it unknown \rm functions $t_{\mu,\nu}$ on
the moduli space $M_{g,N,n_{\al}}$ which are in 1-1 correspondence
with the 1-forms $d\Omega_{\mu,\nu}$. That is, the $t_{\mu,\nu}$
are unknown functions of the coordinates (\ref{eq:modcoord}).
Also, suppose these functions $t_{\mu,\nu}$ form a coordinate
system on a subspace $M^{\prime}$ of $M_{g,N,n_{\al}}$. Let
$\Omega_{\mu,\nu}$ be the abelian integrals of $d\Omega_{\mu,\nu}$
as in (\ref{eq:coordp}). Let $\hat{M}^{\prime}$ be the
tautological bundle over $M^{\prime}$. We now construct a 2-form
$\Pi$ on $\hat{M}^{\prime}$, by using the $t_{\mu,\nu}$ and
$\Omega_{\mu,\nu}$ as follows
\begin{eqnarray}\label{eq:pi}
\Pi=\sum_A \delta \Omega_A \wedge dt_A
\end{eqnarray}
where for simplicity, we write $\left(\mu,\nu\right)=A$ and
\begin{eqnarray*}
\delta \Omega_A =\p_p \Omega_A dp +\sum_B \p_B \Omega_A dt_B
\end{eqnarray*}
When performing the derivative, we treat $t_A$ and $p$ as
independent variables on $\hat{M}^{\prime}$.

Krichever's construction of the Whitham hierarchy is then
equivalent to the following. One wants to ask: how should the
$t_A$ depend on (\ref{eq:modcoord}) if we want $\Pi$ to be simple?
\begin{eqnarray}\label{eq:simple}
\Pi \wedge \Pi=0
\end{eqnarray}

The coefficients of $dp\wedge dt_A\wedge dt_B\wedge dt_C$ of
$\Pi\wedge\Pi$ give the following set of partial differential
equations
\begin{eqnarray*}
\sum\epsilon_{ABC}\p_A\Omega_B\p_p\Omega_C=0
\end{eqnarray*}
where the summation is taken over all the permutations of $A$, $B$
and $C$ and $\epsilon_{ABC}$ is the sign of the permutation. By
taking $C=\left(1,1\right)$, and denote $t_{1,1}$ by $x$, we get
the following
\begin{eqnarray}\label{eq:whitham}
\p_A \Omega_B-\p_B \Omega_A+ \left\{\Omega_A, \Omega_B \right\}=0
\end{eqnarray}
where $\left\{f,g \right\}$ is the Poisson bracket with respect to
the symplectic form $dx \wedge dp$.

This set of partial differential equations is called the truncated
universal Whitham hierarchy $W(n_{\al})$.

By expanding the both sides of these equations in terms of $p$, we
obtain partial differential equations of the coordinates
(\ref{eq:modcoord}) with respect to the $t_A$. By solving these
equations, we obtain the coordinates (\ref{eq:modcoord}) as
functions of $t_A$ and also defines $M^{\prime}$. Of course, the
$t_A$ do not form a complete set of coordinate on the moduli space
$M_{g,N,n_{\al}}$, so the truncated Whitham equations in fact give
the coordinates of the moduli space in terms of the $t_A$ on the
subspace $M^{\prime}$ where the $t_A$ becomes a complete set of
coordinate.

The equations (\ref{eq:whitham}) should be understood as a system
of PDE relating the $t_A$ and the coordinates (\ref{eq:modcoord}),
which enter the equations as the coefficients of the multi-valued
`functions' $\Omega_A$ expanding in terms of the multi-valued
`coordinate' $p$ on the curve $\Gamma_{g,q}$. Namely, we choose
branches of the functions $\Omega_A$ and $p$ such that the
expansions give the coordinates (\ref{eq:modcoord}). The equations
(\ref{eq:whitham}) should only be understood locally. Having this
interpretation in mind, we would now take a closer look of the
consequence on (\ref{eq:whitham}).
\begin{proposition}\label{pro:whitham}
The truncated universal Whitham hierarchy (\ref{eq:whitham}) gives
\begin{eqnarray}\label{eq:whithamcoord1}
\p_AU_i^p&=&\p_xU_i^A, \quad U_i^A=\oint_{b_i}d\Omega_A,\quad i=1,\ldots, g \\
\p_A\sigma_s&=&\p_x\Omega_A\left(\Sigma_s\right),\quad s=1,\ldots,
2g-2\nonumber
\end{eqnarray}
after expanding the terms near $\sigma_s$ and taking the
$b$-periods, where the 1-forms are defined in definition
\ref{de:whithamforms}.
\end{proposition}
Proof. Consider the $b$-periods of the equations
(\ref{eq:whitham}). First note that since $p\mapsto p+U_i^p$ after
going around the cycle $b_i$, the vector fields $\p_A$ becomes
$\p_A\mapsto\p_A+\p_AU_i^p\p_p$ after going around a $b_i$-cycle.
The $b_i$-period of (\ref{eq:whitham}) is therefore
\begin{eqnarray}\label{eq:bperiod}
\left(\p_AU_i^p-\p_xU_i^A\right)\p_p\Omega_B-\left(\p_BU_i^p-\p_xU_i^B\right)\p_p\Omega_A+\p_AU_i^B-\p_BU_i^A
\end{eqnarray}
By multiplying both sides by $dp$, we can view these as equations
of 1-forms. Since the 1-forms $d\Omega_{\mu,\nu}$ are linearly
independent, the coefficient of $d\Omega_B$ must vanish, which
gives the first set of equations in (\ref{eq:whithamcoord}).

The second set of equations is obtained by looking at the behavior
of (\ref{eq:whitham}) near the branch points $\sigma_s$. Near the
branch points $\sigma_s$, we have to use
$\left(p-\sigma_s\right)^{1\over 2}$ as a local coordinate. We
expand the functions $\Omega_A$ near $\sigma_s$ as follows
\begin{eqnarray}\label{eq:nearbranch}
\Omega_A=\Omega_A\left(\Sigma_s\right)+\beta_{A,1}\left(T\right)\left(p-\sigma_s\right)^{1\over
2}+\beta_{A,2}\left(T\right)\left(p-\sigma_s\right)+\cdots
\end{eqnarray}
We now expand (\ref{eq:whitham}) near the point $\sigma_s$. The
term $\p_A\Omega_B$ is of the form
\begin{eqnarray}\label{eq:daomegab}
\p_A\Omega_B=-\p_A\sigma_s\p_p\Omega_B+O\left(1\right)
\end{eqnarray}
and the term $\p_x\Omega_A\p_p\Omega_B$ is of the form
\begin{eqnarray}
\p_x\Omega_A\p_p\Omega_B&=&\p_x\Omega_A\left(\sigma_s\right)\p_p\Omega_B\nonumber\\&+&\beta_{A,1}\left(T\right)\beta_{B,1}\left(T\right)
\left(-{1\over
4}\right)\p_x\sigma_s\left(p-\sigma_s\right)^{-1}\nonumber\\
  &-&\Big({1\over
  2}\beta_{A,1}\left(T\right)\beta_{B,2}\left(T\right)\p_x\sigma_s\\
&+&{1\over
  2}\beta_{A,2}\left(T\right)\beta_{B,1}\left(T\right)\p_x\sigma_s\Big)\left(p-\sigma_s\right)^{-{1\over
  2}}\nonumber\\&+&O\left(1\right) \nonumber
\end{eqnarray}
since the second and the third terms in the right hand side is
symmetric in $A$ and $B$, the expansion of (\ref{eq:whitham}) near
$\sigma_s$ is then
\begin{eqnarray}
\left(-\p_A\sigma_s+\p_x\Omega_A\left(\Sigma_s\right)\right)\p_p\Omega_B+\left(-\p_B\sigma_s+
\p_x\Omega_B\left(\Sigma_s\right)\right)\p_p\Omega_A+
O\left(1\right)
\end{eqnarray}
we now choose $\Omega_B$ such that
$d\Omega_B\left(\Sigma_s\right)\neq 0$. For example, the
holomorphic 1-forms $d\Omega_{h,k}$ only has $2g-2$ zeros while
the number of the points $\Sigma_s$ is $2g$. Therefore there
exists some $B=\left(h,k\right)$ such that
$d\Omega_B\left(\Sigma_s\right)\neq 0$. By similar argument as
before, we arrive at the second set of equations. $\Box$

We also want the truncated universal Whitham hierarchy
$W(n_{\al})$ to be embedded in an infinite hierarchy with
$n_{\al}\rightarrow\infty$. This would give us the following
\begin{proposition}
As $n_{\al}\rightarrow\infty$ for all $\al$, the truncated
universal Whitham hierarchy (\ref{eq:whitham}) gives
\begin{eqnarray}\label{eq:local}
\p_Ak_{\al}-\left\{k_{\al},\Omega_A\right\}\rightarrow 0
\end{eqnarray}
for the 1-forms $\Omega_A$ defined in (\ref{de:whithamforms}).
\end{proposition}
Proof. The proof can be found in \cite{k94}. Let $B=(\al,j)$ and
$A=(\beta,i)$ in (\ref{eq:whitham}). Let $X_-$ be the holomorphic
part of $X$ at $P_{\al}$. Then we have
\begin{eqnarray*}
\p_Ak_{\al}^j-\left\{k_{\al}^j,\Omega_A\right\}=\p_B\Omega_A-\left\{\Omega_A,(\Omega_B)_-\right\}
\end{eqnarray*}
Since $A=(\beta,i)$ and $(\Omega_B)_-$ is holomorphic at
$P_{\al}$, the worst pole that the right hand side could have at
$P_{\al}$ is of order $i$. Therefore
\begin{eqnarray*}
\p_Ak_{\al}^j-\left\{k_{\al}^j,\Omega_A\right\}=O(k_{\al}^{i-j})
\end{eqnarray*}
By letting $j\rightarrow\infty$, the proposition is proved. $\Box$

We can now define the truncated universal Whitham hierarchy as a
system of PDE for the coordinates (\ref{eq:modcoord})
\begin{definition}[Truncated universal Whitham hierarchy]\label{de:whitham}
The truncated universal Whitham hierarchy $W(n_{\al})$ is defined
as the following system of PDE
\begin{eqnarray}\label{eq:whithamcoord}
\p_Ak_{\al}&+& \left\{\Omega_A, k_{\al} \right\}=0   \nonumber \\
\p_AU_i^p&=&\p_xU_i^A, \quad U_i^A=\oint_{b_i}d\Omega_A,\quad i=1,\ldots, g \\
\p_A\sigma_s&=&\p_x\Omega_A\left(\Sigma_s\right),\quad s=1,\ldots,
2g-2\nonumber
\end{eqnarray}
for the 1-forms defined in definition \ref{de:whithamforms}.
\end{definition}
To define these equations, we only use the coefficients of the
$\Omega_A$ up to the $p^{n_{\al}}$ term. Since these coefficients
are expressible in terms of (\ref{eq:modcoord}), the PDE
(\ref{eq:whithamcoord}) defines the functional relation between
the $t_A$ and (\ref{eq:modcoord}).

The set of differential equations (\ref{eq:whitham}) with indices
$A$, $B=\left(\al,i\right)$ only (that is, excluding the indices
$\left(h,k\right)$ in 4.) are called the basic Whitham hierarchy,
they represents the dispersionless limits of integrable
hierarchies.

For simplicity, we would, from now on, call the truncated Whitham
hierarchy $W(n_{\al})$ the Whitham hierarchy.

\paragraph*{Remark} Although the vector fields $\p_A$ are multi-valued on
$\hat{M}_{g,N,n_{\al}}$, they are, nevertheless, well-defined as
normal vectors of the curve
$\Gamma_{g,q}\subset\hat{M}_{g,N,n_{\al}}$. This turns out to be
crucial in our construction of the twistor space.

\subsection{Generating form of the Whitham
hierarchy}\label{sec:genform}

The equations (\ref{eq:whitham}) can be considered as
compatibility conditions of the system
\begin{eqnarray}\label{eq:genE}
\p_{A} E_{\beta}=\left\{E_{\beta}, \Omega_A \right\}
\end{eqnarray}
where $E_{\beta}$ is defined on some open set $V_{\beta}$ on
$\Gamma_{g,q}$. We can perform another change of variable and view
$E_{\beta}, t_A$ as independent variables. The set of equations
(\ref{eq:whitham}) then become
\begin{eqnarray}\label{eq:ewhitham}
\p_A \Omega_B\left(E_{\beta},t\right)=\p_B
\Omega_A\left(E_{\beta},t\right)
\end{eqnarray}
in these new coordinates. The equations (\ref{eq:ewhitham})
suggest the existence of a potential $S\left(E_{\beta},t\right)$
such that
\begin{eqnarray}\label{eq:generating}
\p_A
S_{\beta}\left(E_{\beta},t\right)=\Omega_A\left(E_{\beta},t\right)
\end{eqnarray}
The 2-form $\Pi$ is then
\begin{eqnarray}\label{eq:PiS}
\Pi=\delta E_{\beta} \wedge \delta Q_{\beta}
\end{eqnarray}
where $Q_{\beta}=\p_{E_{\beta}} S_{\beta}$.

The functions $Q_{\beta}$ and $E_{\beta}$ satisfies a `string
equation'
\begin{eqnarray}\label{eq:string}
\left\{Q_{\beta},E_{\beta} \right\}=1
\end{eqnarray}
this implies that $Q_{\beta}$ is also a solution of
(\ref{eq:genE})
\begin{eqnarray}\label{eq:genQ}
\p_A Q_{\beta}=\left\{Q_{\beta}, \Omega_A \right\}
\end{eqnarray}

\section{The twistor space of the Whitham hierarchy (genus-g
curves)}\label{sec: twistorwhit}

We will now proceed to construct the twistor space of the Whitham
hierarchy. We will first construct a twistor space from the
Whitham hierarchy and study its properties. We will then extract
these properties to define a twistor space and recover the Whitham
hierarchy from it. In constructing the twistor space (the direct
construction), we will assume that we are given a solution to the
universal Whitham hierarchy and therefore we know the functional
relation between $t_A$ and the coordinates (\ref{eq:modcoord}) and
also the subspace $M^{\prime}$ of $M_{g,N,n_{\al}}$ where $t_A$
forms a complete set of coordinates. We will denote the
tautological bundle over $M^{\prime}$ by $\hat{M}^{\prime}$. In
particular, the 2-form $\Pi$ is defined on $\hat{M}^{\prime}$.

\subsection{The symplectic reduction of $M^{\prime}$}\label{se:direct}

Following the spirit of \cite{dmt}, we will treat the 2-form $\Pi$
as a presymplectic form on $\hat{M}^{\prime}$ and take the
symplectic reduction of $\left(\hat{M}^{\prime}, \Pi\right)$. This
then give us a 2-dimensional manifold which we will call the
twistor space of the Whitham hierarchy.

We first study the kernel of the 2-form $\Pi$. Due to the
simplicity condition (\ref{eq:simple}), the kernel is generically
of codimension 2. We have the following
\begin{proposition}\label{pro:kernal}
The kernel of the 2-form $\Pi$ is spanned by the vector fields
\begin{eqnarray}\label{eq:kernal}
\p_A-\p_x \Omega_A \p_p +\p_p \Omega_A \p_x
\end{eqnarray}
\end{proposition}
Proof. This can be verified by direct calculation. We first
contract $\p_A$ with $\Pi$, where we use $\left(p,t_A\right)$ as
our independent variables.
\begin{eqnarray*}
\p_A\hook\Pi =\sum_{B\neq (1,1)} \left(\p_A \Omega_B -\p_B
\Omega_A\right)dt_B -\p_p\Omega_A dp-\p_x\Omega_Adx
\end{eqnarray*}
We now apply the (\ref{eq:whitham}) to the above equation and get
\begin{eqnarray}\label{eq:contraction1}
\p_A\hook\Pi =\sum_{B\neq (1,1)} \left(\p_x\Omega_B
\p_p\Omega_A-\p_x\Omega_A\p_p\Omega_B\right)dt_B -\p_p\Omega_A
dp-\p_x\Omega_Adx
\end{eqnarray}
We then consider the contraction between $\Pi$ and $\p_p$, $\p_x$
respectively.
\begin{eqnarray}\label{eq:contraction2}
\p_p\hook\Pi&=&dx+\sum_B\p_p\Omega_Bdt_B \nonumber \\
\p_x\hook\Pi&=&-dp+\sum_B\p_x\Omega_Bdt_B
\end{eqnarray}
By comparing these with (\ref{eq:contraction1}), we see at once
that
\begin{eqnarray*}
\left(\p_A-\p_x\Omega_A\p_p+\p_p\Omega_A\p_x\right)\hook\Pi=0
\end{eqnarray*}
This concludes the proof of the theorem. $\Box$

We can now consider the distribution $Y$ spanned by the vector
fields (\ref{eq:kernal}). Since these span the kernel of a closed
2-form, this distribution is integrable. Note that although the
2-form $\Pi$ is degenerate and singular on some codimension-1
sets, this distribution is still well-defined on these sets. We
would like to consider the leaf space of this distribution. That
is, we would like to study the space $\hat{M}^{\prime}/Y$. This
will be our twistor space ${\mathcal{T}}$.

\paragraph*{Remark} Note that the distribution $Y$ is well-defined
despite the fact that both $\p_A$ and $\Omega_A$ are not
single-valued. The vector field
$\p_A-\p_x\Omega_A\p_p+\p_p\Omega_A\p_x$ is defined up to an
addition of
\begin{eqnarray*}
\p_AU_i^A\p_p-\p_xU_i^p\p_p-\p_xU_i^p\p_p\Omega_A\p_p+\p_p\Omega_A\p_xU_i^A\p_p
\end{eqnarray*}
which vanishes due to the second set of equations in definition
\ref{de:whitham}. The distribution is therefore well-defined.

We can now study the properties of ${\mathcal{T}}$. The following
gives us a more concrete picture of how ${\mathcal{T}}$ looks
like.
\begin{proposition}\label{pro:coord}
The variables $E_{\beta}$, $Q_{\beta}$ and $k_{\al}$ are functions
on ${\mathcal{T}}$. In particular, if we use them as local
coordinates on ${\mathcal{T}}$, the 2-form $\Pi$ then becomes
$\Pi=dE_{\beta} \wedge dQ_{\beta}$ on ${\mathcal{T}}$.

\end{proposition}
Proof. The proof is nothing more than unwinding the definition of
$\left\{f,g\right\}$ in equations (\ref{eq:genE}), (\ref{eq:genQ})
and (\ref{eq:whithamcoord}), which says that given a vector field
$X \in Y$, $X\left(E\right)=X\left(Q\right)=X(k_{\al})=0$. $\Box$

There is also a family of embedded curves in ${\mathcal{T}}$,
which are projections of the curves $\Gamma_{g,q}$. These curves
are labelled by the parameter $t_A$. In explicit terms, these
embeddings are given by
\begin{eqnarray}\label{eq:embedding}
Q_{\beta}=\p_{E_{\beta}} S_{\beta}
\end{eqnarray}
as this equation is defined on the curves $\Gamma_{g,q}$ for
$t_A=constant$.

The 2-form $\Pi$ descends to the twistor space ${\mathcal T}$ to
give a meromorphic section $\Pi$ of the canonical bundle of
${\mathcal T}$, which is non-degenerate except on a codimension-1
subspace. Given a section $T$ of the normal bundle $N_q$ of a
embedded curve $\Gamma_{g,q}\subset {\mathcal{T}}$, the map
$\mu_T=T\hook\Pi|_{\Gamma_{g,q}}$ gives a meromorphic 1-form on
$\Gamma_{g,q}$ which can only have poles at the singular set of
${\mathcal S}$ of $\Pi$ with order less than or equal to that of
$\Pi$. This then defines an isomorphism between
$H^0\left(\Gamma_{g,q},N_q\right)$ and
$H^0\left(\Gamma_{g,q},K\otimes\left[D\right]\right)$ where $K$ is
the canonical bundle of the $\Gamma_{g,q}$ and $\left[D\right]$ is
the line bundle associated to the pole divisor of $\Pi$. Therefore
we have the following
\begin{proposition}\label{pro:normal}
The normal bundle of an embedded curve $\Gamma_{g,q}\in{\mathcal
T}$ is isomorphic to $K \otimes \left[D\right]$, where $K$ is the
canonical bundle of $\Gamma_{g,q}$ and $\left[D\right]$ is the
pole divisor of $\Pi$.
\end{proposition}

In fact, the 1-forms $\p_A\hook\Pi$ are nothing but the 1-forms
$-d\Omega_A$, since we have
\begin{eqnarray*}
\p_A\hook\Pi|_\Sigma&=&\left(\p_x\Omega_A\p_p-\p_p\Omega_A\p_x\right)\hook\left(dE\wedge
dQ\right)|_\Sigma \\
&=&\Big\{\left(\p_xQ\p_p\Omega_A-\p_pQ\p_x\Omega_A\right)\p_pE
\\&-&\left(\p_xE\p_p\Omega_A-\p_pE\p_x\Omega_A\right)\p_pQ \Big\}dp
\end{eqnarray*}
as $dt_A|_\Sigma=0$. We now apply the string equation
(\ref{eq:string}) to obtain
\begin{eqnarray}\label{eq:normal}
\p_A\hook\Pi|_{\Sigma}=\p_p\Omega_Adp=-d\Omega_A
\end{eqnarray}

\subsection{The twistor space of the Whitham
hierarchy}\label{se:twistwhitham}

This concludes our study of the properties of ${\mathcal{T}}$. We
will now define a twistor space independently and show that it
gives a solution to the Whitham hierarchy. We first define the
twistor space of a Whitham hierarchy
\begin{definition}\label{def:twhitham}
A twistor space ${\mathcal{T}}$ of the truncated universal Whitham
hierarchy consists of:

1. A 2-dimensional complex manifold ${\mathcal{T}}$ with a
meromorphic 2-form $\Pi$ which is singular on a divisor $D$,

2. A family of genus-g embedded curves
$\left\{\Sigma_{g,t}\right\}$ and canonical basis of cycles on
each curve,

3. A covering $V_{\beta}$ of a neighborhood $U$ of the singular
divisor $D$, and local coordinates $k_{\beta}^{-1}$ on $V_{\beta}$
such that $k_{\beta}^{-1}$ has zeroes of order 1 at $D$. A fix
point $P_1$ on each curve $\Sigma_t$ such that $P_1\in D$.
\end{definition}
The main result in this section is the following
\begin{theorem}\label{thm:twistorwhitham}
There is a one-one correspondence between a solution of the
Whitham hierarchy and a twistor space ${\mathcal{T}}$ defined
above.
\end{theorem}
Proof. Given a Whitham hierarchy, we can take the symplectic
reduction of its moduli space as we did in the last section to get
a twistor space ${\mathcal{T}}$. The local coordinates $k_{\al}$
near each marked point $P_{\al}$ is well-defined on the twistor
space due to (\ref{eq:local}) and they give the local coordinates
in 3. of definition \ref{def:twhitham}.

This gives the first half of the proof. To go the other way round,
we want to recover the space $\hat{M}^{\prime}$ and show that the
pull-back $\tilde{\Pi}$ of the 2-form $\Pi$ to $\hat{M}^{\prime}$
has the correct form. The simplicity condition follows
automatically as $\tilde{\Pi}$ is the pull-back of a 2-form on a
2-dimensional manifold.

The space $M^{\prime}$ is the set of curves $\left\{
\Sigma_{g,t}\right\}$. By considering $E$ and $Q$ as functions of
the $t_A$, we can pull back the 2-form $\Pi$ to the space
$\hat{M}_{g,N}$. This can be achieved by expanding $E$ and $Q$
near the the divisor $D$ in terms of the local coordinate
\begin{eqnarray*}
p=-\int d\Omega_{1,1}
\end{eqnarray*}
and consider the coefficients as functions of $t$.

Now choose representatives $\p_A$ of normal vectors such that
\begin{eqnarray}\label{eq:reconstructta}
\p_A\hook\Pi=\left(\p_x\Omega_A\p_p-\p_p\Omega_A\p_x\right)\hook\Pi
\end{eqnarray}
where $\p_x\hook\Pi=-dp$. We see that
\begin{eqnarray*}
\left(\p_x\Omega_A\p_p-\p_p\Omega_A\p_x\right)\hook\Pi|_{\Sigma}=\p_p\Omega_Adp
\end{eqnarray*}
since $\mu_{\p_x}=-dp$. We see that $t_A$ is a complete set of
coordinates on $M^{\prime}$ because the $\p_A$ generates all the
possible deformations of the curves and are independent.

We have
\begin{eqnarray}\label{eq:pullbackpi}
\tilde{\Pi}=\left(\sum_A\p_AEdt_A+\p_pEdp\right)\wedge\left(\sum_B\p_BQdt_B+\p_pQdp\right)
\end{eqnarray}
We now contract $\tilde{\Pi}$ with $\p_A$ and make use of
(\ref{eq:reconstructta}). The $dt_B$ component of
$\p_A\hook\tilde{\Pi}$ then gives
\begin{eqnarray}\label{eq:xbcom}
\p_x\Omega\p_pE\p_BQ-\p_pQ\p_x\Omega_A\p_BE+\p_p\Omega_A\p_xE\p_BQ-\p_p\Omega_A\p_xQ\p_BE
\end{eqnarray}
while the $dp$ component gives
\begin{eqnarray}\label{eq:pcom}
\p_p\Omega_A\p_xE\p_pQ-\p_p\Omega_A\p_xQ\p_pE
\end{eqnarray}
since this is just $-\p_p\Omega_A$, we obtain the string equation
(\ref{eq:string}). Note that (\ref{eq:reconstructta}) can be
written as
\begin{eqnarray*}
\left(\p_A-\p_x\Omega_A\p_p+\p_p\Omega_A\p_x\right)\left(E\right)dQ=
\left(\p_A-\p_x\Omega_A\p_p+\p_p\Omega_A\p_x\right)\left(Q\right)dE
\end{eqnarray*}
this gives (\ref{eq:genE}) and (\ref{eq:genQ}), which implies the
compatibility of the flows (\ref{eq:whitham}). By applying
(\ref{eq:genE}) and (\ref{eq:genQ}) to (\ref{eq:xbcom}) and making
use of the string equation (\ref{eq:string}), the expression
(\ref{eq:xbcom}) becomes
\begin{eqnarray}
\p_x\Omega_A\p_p\Omega_B-\p_p\Omega_A\p_x\Omega_B=\p_B\Omega_A-\p_A\Omega_B
\end{eqnarray}
this proves theorem follows. $\Box$

\paragraph*{Remark} As in the case of dKP equation \cite{dmt}, the Whitham hierarchy
depends on the choice of coordinates, which is reflected in 3. of
definition \ref{def:twhitham}.

\section{Examples}\label{se:exwhitham} We will now look at a few
examples of twistor spaces of Whitham hierarchies.
\begin{example}\label{ex:dKP}
{The dKP equation}\cite{dmt}
\end{example}
The first example is the dKP equation, its twistor space was first
constructed in \cite{dmt} from the Einstein-Weyl metric. This is
the case of a Whitham hierarchy with genus-0 curves.

The twistor space of this equation is given by a 2-dimensional
complex manifold ${\mathcal{T}}$ such that the 2-form $\Pi$ is
singular at a connected set $D$ with order 4. The Darboux
coordinates $E$, $Q$ are chosen so that $E$, $Q$ are singular on
$D$ with orders 1 and 2 respectively. There is only one fixed
point $E=Q=\infty$. The local coordinate is chosen to be $Q$. The
normal bundles of the curves are then
${\mathcal{O}}\left(2\right)$. (since the order of poles of the
normal bundle at $p=\infty$ cannot exceed the correspondent one of
$Q$).

The solution to the Whitham hierarchy is recovered by expanding
$E$ and $Q$ near $z=\infty$.
\begin{eqnarray}
Q=p+\sum u_i p^{-1}, \quad E=\sum v_iQ^{-i}+x+Qy+Q^2t
\end{eqnarray}
By using $\Pi=dE\wedge dQ$ and the singular structure of $\Pi$, we
see that the pull-back $\tilde{\Pi}$ has the form
\begin{eqnarray}
\tilde{\Pi}=dx\wedge dp+dy\wedge d\left({1\over
{2}}p^2+u_1\right)+dt\wedge d\left({1 \over 3}p^3+pu_1+w_1\right)
\end{eqnarray}
the simplicity of this 2-form then gives a solution to the dKP
equation
\begin{eqnarray}
\left(\left(u_1\right)_t-u_1
\left(u_1\right)_x\right)_x=\left(u_1\right)_{yy}
\end{eqnarray}

\begin{example}{Elliptic curves}\label{ex:elliptic}
\end{example}
We now look at the set of elliptic curves $\Gamma_q$ in
$\mathbb{CP}^1\times\mathbb{CP}^1$, these are given the equation
\begin{eqnarray}\label{eq:elliptic}
w^2=4z^3-g_1z-g_2
\end{eqnarray}
We take the twistor space to be $\mathbb{CP}^1\times\mathbb{CP}^1$
with coordinates $\left(w,z\right)$ and consider the deformations
of the curve.

We first choose the 2-form to be $dw\wedge dz$, and we will choose
the local coordinate $k$ to be ${w\over z}$. The marked point will
be the point where $dw\wedge dz$ be comes singular on the curve,
that is, the point $\left(\infty,\infty\right)$. We shall simply
denote this point by $\infty$. Let $\lambda$ be the uniformization
parameter of the curve, and $\wp\left(\lambda\right)$ be the
Weierstrass function. We choose our basis of cycle such that
\begin{eqnarray}\label{eq:acycle}
\oint_a d\lambda&=&{{2g_3^{-{1\over 10}}g_7^{-{1\over 15}}}\over
4}
\end{eqnarray}
Since $z=\wp\left(\lambda\right)$ and
$w=\wp^{\prime}\left(\lambda\right)$, we see that the divisor $D$
is of order 7 on the curve. The space
$H^0\left(\Gamma_q,N_q\right)$ is therefore of dimension 7 (since
there is no meromorphic 1-form with a simple pole on $\Gamma_q$,
the dimension is 7 instead of 8). We also want to fix the marked
point $\left(\infty,\infty\right)$ so we want all the curves to
intersect this point. This means we only consider a 6-dimensional
subspace of $H^0\left(\Gamma_q,N_q\right)$. In particular, since
the normal vector $N$ such that $\mu_N$ has a pole of order 7 does
not vanish at $\left(\infty,\infty\right)$, it will move the
marked point and we shall not consider flows generated by such
normal vectors.

We shall see in a moment that the full family of curves obtained
from deforming (\ref{eq:elliptic}) is given by
\begin{eqnarray}\label{eq:fullelliptic}
g_3w^2+g_4w+g_5wz+=g_6z^3+g_7z^2-g_1z-g_2
\end{eqnarray}
subject to a scale invariant such that
\begin{eqnarray}\label{eq:scale}
\sum g_i\p_{g_i}=0
\end{eqnarray}

Each curve in (\ref{eq:fullelliptic}) can be brought to the
standard form (\ref{eq:elliptic}) by a linear transformation
\begin{eqnarray}\label{eq:trans}
w^{\prime}&=&aw+bz+c ,\quad z^{\prime}=ez+f \nonumber\\
(w^{\prime})^2&=&4(z^{\prime})^3-\tilde{g}_1z^{\prime}-\tilde{g}_2
\end{eqnarray}
therefore we can write the Weierstrass functions as
\begin{eqnarray}\label{eq:weierstrass}
\wp(\lambda)=ez+f,\quad \wp^{\prime}(\lambda)=aw+bz+c
\end{eqnarray}
By comparing (\ref{eq:trans}) with (\ref{eq:fullelliptic}), we see
that the coefficients $a,\ldots,f$ are given by
\begin{eqnarray}\label{eq:transcoef}
a&=&g_3^{1\over 2},\quad b={1\over 2}g_5g_3^{-{1\over 2}},\quad
c={1\over 2}g_4g_3^{-{1\over 2}}, \nonumber \\
e&=&\Big({g_7\over 4}\Big)^{-{1\over 3}},\quad f={1\over
{12e^2}}\Big(g_8+{{1\over 4}g_5^2g_3^{-1}}\Big) \\
\tilde{g}_1&=&{1\over e}\Big(g_1+12ef^2-{1\over
4}g_4g_5g_3^{-1}\Big)
\nonumber \\
\tilde{g}_2&=&g_2-\tilde{g}_1f+4f^3-c^2\nonumber
\end{eqnarray}
We can therefore compute various powers of the jet $k$ in terms of
the Weierstrass function and hence compute the 1-forms
$d\Omega_{\infty,i}$. We shall now denote $d\Omega_{\infty,i}$ by
$d\Omega_i$.
\begin{eqnarray}\label{eq:kjets}
d\Omega_1&=&{{2e}\over a}(\wp(\lambda)-\omega)d\lambda \nonumber \\
d\Omega_2&=&4\Big({e\over
a}\Big)^2\Big(\wp^{\prime}(\lambda)-{b\over
e}(\wp(\lambda)-\omega)\Big)d\lambda \nonumber \\
d\Omega_3&=&\Big({e\over
a}\Big)^3\Bigg[4\wp^{\prime\prime}(\lambda)-{{12b}\over
e}\wp^{\prime}(\lambda)+\Bigg(24f+6\Big({b\over
e}\Big)^2\Bigg)(\wp(\lambda)-\omega)\Bigg]d\lambda \nonumber \\
d\Omega_4&=&\Big({e\over
a}\Big)^4\Bigg[32\wp^{\prime}(\lambda)\wp(\lambda)-{{16b}\over
e}\wp^{\prime\prime}(\lambda)  \\
&+&\Bigg(24\Big({b\over e}\Big)^2+64f\Bigg)\wp^{\prime}(\lambda) \nonumber \\
&+&\Bigg(-96{{bf}\over e}-32c-8\Big({b\over
e}\Big)^3\Bigg)(\wp(\lambda)-\omega)\Bigg]
\nonumber \\
d\Omega_5&=&\Big({e\over
a}\Big)^5\Bigg[16(\wp^{\prime}(\lambda))^2+16\wp(\lambda)\wp^{\prime\prime}(\lambda)-160{b\over
e}\wp^{\prime}(\lambda)\wp(\lambda) \nonumber \\
&+&\Bigg(40\Big({b\over
e}\Big)^2+80f\Bigg)\wp^{\prime\prime}(\lambda)+\Bigg(-320{{bf}\over
e}-80c-40\Big({b\over e}\Big)^3\Bigg)\wp^{\prime}(\lambda)
\nonumber \\
&+&\Bigg(240\Big({b\over e}\Big)^2f+160{{bc}\over
e}-16\tilde{g}_1+10\Big({b\over
e}\Big)^4+480f^2\Bigg)(\wp(\lambda)-\omega)\Bigg]
\end{eqnarray}
where $\omega$ is the the a-period of $\wp(\lambda)d\lambda$
divided by the factor ${{2g_3^{-{1\over 10}}g_7^{-{1\over
15}}}\over 4}$ in (\ref{eq:acycle}). The holomorphic 1-form
$d\Omega_{h,1}$, which we will denote by $d\Omega_0$ is
\begin{eqnarray}\label{eq:omega0}
d\Omega_0=\Bigg({{2g_3^{-{1\over 10}}g_7^{-{1\over 15}}}\over
4}\Bigg)^{-1}d\lambda
\end{eqnarray}

The Whitham hierarchy that we are trying to solve is the form
\begin{eqnarray}\label{eq:whithamelliptic}
\p_jd\Omega_i-\p_id\Omega_j+ \left\{d\Omega_i,d\Omega_j\right\}=0
\end{eqnarray}
The coefficients of the expansion of the above equations can be
expressed in terms of the $g_i$ in (\ref{eq:fullelliptic}). We
will therefore treat the coefficients $g_i$ as a solution to
(\ref{eq:whithamelliptic}) and express them in terms of $t_i$. In
fact, we will express $t_i$ in terms of the $g_i$.

We first look at the tangent bundle of the curve $\Gamma_q$. By
differentiating (\ref{eq:fullelliptic}), we see that the tangent
bundle is spanned by
\begin{eqnarray}\label{eq:tangent}
\Im=(2g_3w+g_4+g_5z)\p_z-(3z^2g_6+2zg_7-g_1-g_5w)\p_w
\end{eqnarray}
and the normal bundle of the curve $\Gamma_q$ is spanned by
\begin{eqnarray}\label{eq:normalbundle}
\left\{w\p_w-z\p_z,\p_w,z\p_w,\p_z,{{\p_w}\over
{(2g_3w+g_4+g_5z)}},{{z\p_w}\over {(2g_3w+g_4+g_5z)}}\right\}
\end{eqnarray}
since the contraction of these vector fields with $dw\wedge dz$
gives 1-forms of desired order.

We can also see this as follows. Since the poles of $dw$, $dz$ are
of order 4 and 3 at $\infty$, $\p_w$ and $\p_z$ vanish to the
order 4 and 3 at $\infty$ respectively. Therefore the vector
fields $w\p_w$ etc. are holomorphic on $\Gamma_q$. However, the
vector field $w^2\p_w$ and $z^2\p_z$ which are holomorphic on
$\mathbb{CP}^1\times\mathbb{CP}^1$ are not holomorphic on
$\Gamma_q$. On the other hand, vector fields like $z\p_w$ that are
not holomorphic on $\mathbb{CP}^1\times\mathbb{CP}^1$ becomes
holomorphic on $\Gamma_q$ as $\Gamma_q$ only intersects $z=\infty$
and $w=\infty$ at $(\infty,\infty)$. Also, the vector fields
${{\p_w}\over {(2g_3w+g_4+g_5z)}}$ and ${{z\p_w}\over
{(2g_3w+g_4+g_5z)}}$ are holomorphic as $\p_w$ is tangent to
$\Gamma_q$ when the denomination vanishes on $\Gamma_q$, which we
can see by setting $\Im=0$ in (\ref{eq:tangent}). Therefore the
above are holomorphic normal bundles on $\Gamma_q$.

Since $\p_w$ acts on $\mathbb{CP}^1\times\mathbb{CP}^1$ by an
infinitesmal translation of $w$, we see that the normal vectors
(\ref{eq:normalbundle}) generates the following movements of the
curve.
\begin{eqnarray*}
w\p_w-z\p_z&=&5g_3\p_{g_3}+2g_4\p_{g_4}+3g_2\p_{g_2}\nonumber \\
&&+4g_1\p_{g_1}+g_7\p_{g_7}+3g_5\p_{g_5} \nonumber \\
 \p_w&=&-2g_3\p_{g_4}+g_4\p_{g_2}+g_5\p_{g_1},\nonumber \\
z\p_w&=&2g_3\p_{g_5}+g_4\p_{g_1}-g_5\p_{g_7},  \nonumber \\
\p_z&=&g_5\p_{g_4}+3g_6\p_{g_7}-2g_7\p_{g_1}+g_1\p_{g_2},\\
&{{\p_w}\over
{(2g_3w+g_4+g_5z)}}&=\p_{g_2},\nonumber \\
 &{{z\p_w}\over
{(2g_3w+g_4+g_5z)}}&=\p_{g_1} \nonumber
\end{eqnarray*}
where we have used the scale invariance $\sum_ig_i\p_{g_i}=0$ to
eliminate the $g_6\p_{g_6}$ term in the first equation. We see
that the full family of curves is of the form
(\ref{eq:fullelliptic}). By contracting the normal vectors with
respect to the 2-form $\Pi$ and using $\mu_{t_i}=-d\Omega_i$, we
see that
\begin{eqnarray}\label{eq:couple}
\p_{g_1}&=&{1\over g_6}\p_{t_1}+\p_{g_1}t_0\p_{t_0} \nonumber \\
\p_{g_2}&=&g_3^{-{3\over 5}}g_6^{-{2\over 5}}\p_{t_0} \nonumber \\
\p_{g_3}&=&{g_3\over {5g_6}}\p_{t_5}+{1\over 4}{g_5\over
g_6^2}\p_{t_4}-{g_8\over {3g_6^2}}\p_{t_3}+\p_{g_3}t_0\p_{t_0} \nonumber \\
\p_{g_4}&=&{1\over{2g_6}}\p_{t_2}+\p_{g_4}t_0\p_{t_0} \\
\p_{g_5}&=&{g_3\over{4g_6^2}}\p_{t_4}+{g_5\over{3g_6^2}}\p_{t_3}-{g_7\over{2g_6^2}}\p_{t_2}
+\p_{g_5}t_0\p_{t_0} \nonumber \\
\p_{g_7}&=&-{g_3\over{3g_6^2}}\p_{t_3}-{g_5\over{2g_6^2}}\p_{t_2}+{g_7\over
g_6^2}\p_{t_1}+\p_{g_5}t_0\p_{t_0} \nonumber
\end{eqnarray}
while the functions $\p_{g_i}t_0$ are expressed in terms of $g_i$
and $\omega$. From (\ref{eq:couple}), we see that the functions
\begin{eqnarray}\label{eq:sol}
t_1&=&{{2g_1+g_7^2}\over{2g_6}}+c_1 \nonumber \\
t_2&=&{{g_4-g_5g_7}\over{2g_6}}+c_2\nonumber \\
t_3&=&-{{g_8g_3+g_5^2}\over{3g_6^2}}+c_3 \\
t_4&=&{{g_3g_5}\over{4g_6^2}}+c_4\nonumber \\
t_5&=&{g_3^2\over{5g_6}}+c_5 \nonumber
\end{eqnarray}
where $c_i$ are integration constants. While $t_0$ satisfies the
following equations
\begin{eqnarray}\label{eq:t0der}
\p_{g_1}t_0&=&g_3^{-{3\over 5}}g_6^{-{2\over 5}}\Bigg[\omega
e^{-1}-{1\over3}\Bigg({g_7\over
{g_6}}+{{g_5^2}\over{4g_6g_3}}\Bigg)\Bigg]
\nonumber \\
\p_{g_2}t_0&=&g_3^{-{3\over 5}}g_6^{-{2\over 5}} \nonumber \\
\p_{g_3}t_0&=&{1\over{5g_3}}\Bigg[\Bigg({{g_4^2}\over
g_3}-3g_2-{{g_7g_5g_4}\over{6g_6g_3}}+{{g_7g_1}\over{3g_6}}-{{g_5^3g_4}\over{4g_3^2g_6}}
\nonumber \\
&+&{{g_5^2g_1}\over{2g_3g_6}}\Bigg)\p_{g_2}t_0+\Bigg({{g_4g_5}\over g_3}-4g_1 \nonumber \\
&-&{{g_7g_5^2}\over{6g_6g_3}}-{{2g_7^2}\over{3g_6}}-{{g_5^4}\over{4g_3^2g_6}}-{{g_5^2g_7}\over{g_6g_3}}
+{{3g_5g_4}\over{2g_3}}\Bigg)\p_{g_1}t_0\Bigg]  \\
\p_{g_4}t_0&=&{1\over{2g_3}}\Big(-g_4\p_{g_1}t_0-g_5\p_{g_2}t_0\Big)
\nonumber \\
\p_{g_5}t_0&=&{1\over{2g_3}}\Bigg[\Bigg({{g_5^2g_4}\over{6g_6g_3}}-{{g_5g_1}\over{3g_6}}\Bigg)\p_{g_2}t_0
\nonumber \\
&+&\Bigg({{g_3^2}\over{6g_6g_3}}+{{2g_5g_7}\over{3g_6}}-g_4\Bigg)\p_{g_1}t_0\Bigg]
\nonumber \\
\p_{g_7}t_0&=&{1\over{3g_6}}\Bigg[\Bigg({{g_5g_4}\over{2g_3}}-g_1\Bigg)\p_{g_2}t_0+\Bigg({{g_5^2}\over{2g_3}}+2g_7\Bigg)
\p_{g_1}t_0\Bigg] \nonumber
\end{eqnarray}
where $e=(4^{-1}g_6)^{1\over3}$.

This gives a solution to the Whitham hierarchy
(\ref{eq:whithamelliptic}) given by twistor theory.

\begin{example}\label{ex:algcurves} {Algebraic
curves in $\mathbb{CP}^1 \times \mathbb{CP}^1$}
\end{example}
This example generalized the last example to the case where the
curves are given by a polynomial
\begin{eqnarray*}
P\left(w,z\right)=0
\end{eqnarray*}
In this case we choose the symplectic form to be $dz\wedge dw$.
The twistor space is then a subset of the normal bundle which
consists of normal vectors $T$ such that $\mu_T|_\Sigma$ are
normalized 1-forms. This gives us the deformations of the curve
$\Sigma$ that correspond to the Whitham deformations.

\paragraph*{Spectral curves of the dual isomonodromic deformations}

The isomonodromic problem was studied by Jimbo, Miwa and Ueno.
\cite{jmu81a}, \cite{jmu81b} Suppose we have an linear system of
ODE
\begin{eqnarray}\label{eq:iso}
{d\over dz}\Psi(z)=A(z)\Psi (z)
\end{eqnarray}
where $A(z)$ is a $n\times n$ matrix-valued rational function in
$z$
\begin{eqnarray*}
A(z)=\sum_{\al\in{\mathcal{D}}}{A_{\al}(z)\over
(z-\al)^{r_{\al}}}+A_{\infty}(z)
\end{eqnarray*}
where $A_{\al}(z)$ are polynomials in $z-\al$ of degrees
$r_{\al}-1$, $A_{\infty}(z)$ is a polynomial of degree
$r_{\infty}$ and ${\mathcal{D}}$ is a divisor in $\mathbb{CP}^1$.
We call the system non-resonant if the leading terms of $A_{\al}$
has distinct eigenvalues when $r_{\al}>1$ and that no eigenvalues
of the leading terms of $A_{\al}$ differ by an integer when
$r_{\al}=1$. In this case, we can find unique formal solutions to
the system of the form
\begin{eqnarray}\label{eq:formal}
\Psi(z-\al)=Y^{(\al)}(z)\exp\left(D_{\al}(z)+m_{\al}\log (z-\al
)\right)
\end{eqnarray}
as $z\rightarrow\al$, where
$Y^{(\al)}(z)=\sum_{j=0}^{\infty}Y_j^{(\al)}(z-\al)^{-j}$ is a
formal Taylor series of matrix that is invertible at $z=\al$ and
$D_{\al}(z)$ is a diagonal matrix with polynomial entries in
$z-\al$. The matrix $m_{\al}$ is a diagonal matrix and is called
the exponent of formal monodromy.

It is known (e.g. \cite{w76}) that near each pole $\al$ of $A(z)$,
there exist sectors $S_{\al}^{(j)}$ on which there exist solutions
$\Psi_{\al}^{(j)}$ that are asymptotic to the one in
(\ref{eq:formal}).

By comparing solutions on different sectors, one obtains the
Stokes matrices
$S_{jk}^{(\al)}=\left(\Psi_{\al}^{(j)}\right)^{-1}\Psi_{\al}^{(k)}$.
Similarly, analytic continuation of solutions near different poles
define the connection matrices. \cite{jmms} Let us denote the
coefficients of $D_{\al}$ and the pole positions $\al$ by $T$
collectively. The isomonodromic problem is to find the dependence
of the matrix $A(z)$ on $T$ such that the Stokes matrices, the
connection matrices and the exponent of formal monodromy remain
constant. It was shown in \cite{jmu81a} that the dependence
$A(z,T)$ of $A(z)$ on $T$ is determined by the following
differential equation of $A(z,T)$
\begin{eqnarray}\label{eq:Azc}
{\rm d}A(z,T)=\p_zB(z,T)-\left[A(z,T),B(z,T)\right]
\end{eqnarray}
where ${\rm d}$ is the exterior derivative with respect to $T$ and
$B(z,T)$ is a rational matrix 1-form.

The `spectral curve' of the isomonodromic problem is then defined
by the algebraic equation \cite{h94}, \cite{sw}
\begin{eqnarray}\label{eq:spec}
\det\left(w -A(z,T)\right)=0
\end{eqnarray}
Unlike the spectral curves of other integrable systems, the
`spectral curve' of an isomonodromic problem is not preserved by
the isomonodromic flows. One can see this from (\ref{eq:Azc}). The
term $\left[A(z,T),B(z,T)\right]$ on the right hand side of
(\ref{eq:Azc}) would preserve the spectral invariance whereas the
term $\p_zB(z,T)$ would lead to deformations of the spectral
curve.

From now on we will consider a simple case in which the matrix
$A(z)$ is of the following form
\begin{eqnarray}\label{eq:isork1}
A(z)=\sum_{i=1}^r{A_{i}\over (z-\al_i)}+C
\end{eqnarray}
where $A_{i}$ are rank 1 matrices constant in $z$ and $C$ is a
diagonal matrix constant in $z$, $C=\diag(c_1,\ldots,c_n)$. The
deformation parameters in this case are the $\al_i$ and the $c_i$.

In this case, a technique called `duality' \cite{h94} can be
employed to obtain an algebraic curve that is birational to the
spectral curve. Moreover, this curve is defined by the zero locus
of a polynomial $P(z,w)=0$ in $\mathbb{CP}^1\times\mathbb{CP}^1$.
These curves were considered by Sanguinetti and Woodhouse.
\cite{sw} The explicit construction goes as follows. Let $\Delta$
be the diagonal matrix $\Delta=\diag(\al_1,\ldots,\al_r)$. Since
all the $A_i$ are of rank 1, we can write the matrix $A(z)$ as
follows
\begin{eqnarray}\label{eq:factor}
A(z)=G^T(\Delta-z)^{-1}F+C
\end{eqnarray}
where $G$, $F$ are $n\times r$ matrices constant in $z$.

There is a `dual isomondromic problem' associated to it, which is
defined by
\begin{eqnarray}\label{eq:dual}
\p_w-F(c-w)^{-1}G^T-\Delta
\end{eqnarray}
It was shown \cite{h94} that when $A(z)$ is deformed
isomondromically, the dual system also undergoes isomonodromic
deformation.

The \it dual isomonodromic spectral curve \rm is defined by the
following polynomial in $w$ and $z$
\begin{eqnarray}\label{eq:spectral}
\det \left[ \mathbb{M}- \pmatrix{
  z & 0 \cr
  0 & w\cr} \right]=0
\end{eqnarray}
where $\mathbb{M}$ is the following matrix
\begin{eqnarray*}
\mathbb{M}=\pmatrix{
  \Delta & -F \cr
  G^t & C\cr}
\end{eqnarray*}
The genus of such curve is shown in \cite{sw} to be $(n-1)(r-1)$.
The whole set of curves $\Sigma_T$ obtained by deforming the
spectral curve through isomonodromic deformations form a $n+r$
parameter family of curves. This is a subset of curves defined by
the zero locus of all polynomials
\begin{eqnarray*}
\sum_{i=0}^r\sum_{j=0}^nX_{ij}z^iw^j=0
\end{eqnarray*}
which is a $nr=2n+2r+g$ parameter family of curves. We can treat
this family of curves and the 2-dimensional complex manifold
$\mathbb{CP}^1\times\mathbb{CP}^1$ as a twistor space of Whitham
hierarchy as in example \ref{ex:algcurves}. The 2-form is given by
$dz\wedge dw$ and the fixed points are the intersection points
between $z=\infty$, $w=\infty$ and the curves. These are the
points $(z=\infty, w=c_i)$ and $(z=\al_i,w=\infty)$. The local
coordinates near $z=\infty$ are chosen to be $z^{-1}$ and the
local coordinates near $w=\infty$ are chosen to be $w^{-1}$. In
this case there is no natural choice of the point $P_1$ and we
will choose it to be $(z=\al_1,w=\infty)$.

In this case the Whitham times can be solved by the hodograph
method. \cite{k94}, \cite{ts} They are given by
\begin{eqnarray}\label{eq:whittime}
t_{\al_i,j}&=&{\rm res}_{w=\infty}w^{-j}zdw, \quad
j=0,1 \nonumber\\
t_{c_i,j}&=&-{\rm res}_{z=\infty}z^{-j}wdz, \quad
j=0,1 \\
t_{h,j}&=&\oint_{a_j}zdw,\quad j=1,\ldots,g \nonumber
\end{eqnarray}
where we used the index $\al_i$ to denote the point
$(z=\al_i,w=\infty)$ and the index $c_i$ to denote the point
$(z=\infty,w=c_i)$. The residues are taken around the
corresponding points. The integrals defining the $t_{h,j}$ are
around a certain choice of $a$-cycles on the curve. Note that
these formula are independent on the choice of $P_1$.

Since the Whitham flows generates all the deformations of the
algebraic curve, we can express the isomonodromic deformations of
the spectral curve in terms of the Whitham flows.
\begin{eqnarray}\label{eq:isovwhit}
{\p\over\p {\al_i}}&=&\sum_A{{\p t_A}\over {\p {\al_i}}}\p_A
\nonumber \\
{\p\over\p {c_i}}&=&\sum_A{{\p t_A}\over {\p {c_i}}}\p_A
\end{eqnarray}
To see how the isomonodromic times are related to the Whitham
times, we first consider the behavior of $w$ near
$(z=\infty,w=c_i)$. Near these points, $w$ has the Laurent
expansion
\begin{eqnarray}\label{eq:inflaurent}
w=c_i+m_{\infty}^iz^{-1}+O(z^{-2})
\end{eqnarray}
where $m_{\infty}^i$ is the $i^{th}$ entry in the formal exponent
of monodromy at $z=\infty$. From (\ref{eq:whittime}) we see that
\begin{eqnarray}\label{eq:whittimew}
t_{c_i,0}=-m_{\infty}^i, \quad t_{c_i,1}=-c_i
\end{eqnarray}
Similarly, near $(z=\al_i,w=\infty)$, $z$ has the expansion
\begin{eqnarray}\label{eq:allaurent}
z=\al_i+n_{\infty}^iw^{-1}+O(w^{-2})
\end{eqnarray}
where $n_{\infty}^i$ is the $i^{th}$ entry in the formal exponent
of monodromy at $w=\infty$ of the dual system (\ref{eq:dual}). We
see that
\begin{eqnarray}\label{eq:whittimez}
t_{\al_i,0}=n_{\infty}^i, \quad t_{\al_i,1}=\al_i
\end{eqnarray}
By substituting these into (\ref{eq:isovwhit}) we obtain
\begin{eqnarray}\label{eq:isovwhit2}
{\p\over\p {\al_i}}&=&\p_{\al_i,1}+\sum_{k=1}^g{{\p }\over {\p
{\al_i}}}\left(\oint_{a_k}zdw\right)\p_{h,k}
\nonumber \\
{\p\over\p {c_i}}&=&\p_{c_i,1}+\sum_{k=1}^g{{\p }\over {\p
{c_i}}}\left(\oint_{a_k}zdw\right)\p_{h,k}
\end{eqnarray}
To calculate the second terms, first note that since $z$ is single
value on the spectral curve, the derivatives ${{\p }\over {\p
{c_i}}}\left(\oint_{a_k}zdw\right)$ and ${{\p }\over {\p
{\al_i}}}\left(\oint_{a_k}zdw\right)$ has no boundary terms. That
is, we have
\begin{eqnarray}\label{eq:deriv}
{{\p }\over {\p
{c_i}}}\left(\oint_{a_k}zdw\right)&=&\oint_{a_k}{{\p }\over {\p
{c_i}}}zdw \\
{{\p }\over {\p
{\al_i}}}\left(\oint_{a_k}zdw\right)&=&\oint_{a_k}{{\p }\over {\p
{\al_i}}}zdw
\end{eqnarray}
To evaluate $\p_{c_i}zdw$ or $\p_{\al_i}zdw$, we can either think
of the isomonodromic flows as flows that keep $z$ fixes and
changes $w$ as in the original isomonodromic system
(\ref{eq:iso}), or we could think of these flows as keeping $w$
fix and changing $z$ as in the dual isomonodromic system
(\ref{eq:dual}). To see that these two different interpretations
make no difference in the final result, let us denote the
determinant in (\ref{eq:spectral}) by $P(z,w,T)$. Let
\begin{eqnarray*}
P(z,w,T)=\sum_{i=0}^r\sum_{j=0}^nX_{ij}(T)z^iw^j=0
\end{eqnarray*}
Then we see that
\begin{eqnarray*}
\p_{\al_i}z&=&-\left(\sum_{i=0}^r\sum_{j=0}^n\p_{\al_i}X_{ij}(T)z^iw^j\right)\left(\p_zP(z,w,T)\right)^{-1}
\\
\p_{\al_i}w&=&-\left(\sum_{i=0}^r\sum_{j=0}^n\p_{\al_i}X_{ij}(T)z^iw^j\right)\left(\p_wP(z,w,T)\right)^{-1}
\end{eqnarray*}
and similar expressions for $\p_{c_i}w$ and $\p_{c_i}z$. If we
think of the isomonodromic flows as flows that fix $w$, the
integrals in (\ref{eq:deriv}) are then
\begin{eqnarray}\label{eq:wfix}
\oint_{a_k}{{\p }\over {\p
{\al_i}}}zdw=-\oint_{a_k}\left(\sum_{i=0}^r\sum_{j=0}^n\p_{\al_i}X_{ij}(T)z^iw^j\right)\left(\p_zP(z,w,T)\right)^{-1}dw
\end{eqnarray}
on the other hand, if we think of the flows as flows that fix $z$,
the integrals become
\begin{eqnarray}\label{eq:zfix}
\oint_{a_k}{{\p }\over {\p
{\al_i}}}zdw=\oint_{a_k}\left(\sum_{i=0}^r\sum_{j=0}^n\p_{\al_i}X_{ij}(T)z^iw^j\right)\left(\p_wP(z,w,T)\right)^{-1}dz
\end{eqnarray}
since $\oint_{a_k}zdw=-\oint_{a_k}wdz$ as $zdw+wdz$ is a total
differential.

To see that the right hand sides of (\ref{eq:wfix}) and
(\ref{eq:zfix}) are the same, note that by differentiating the
expression $P(z,w,T)=0$, we have
$\left(\p_wP(z,w,T)\right)^{-1}dz=-\left(\p_zP(z,w,T)\right)^{-1}dw$.
By replacing the $\al_i$ derivatives in (\ref{eq:wfix}) and
(\ref{eq:zfix}) by derivatives of $c_i$ and apply similar
argument, we see that the same holds for the $c_i$ derivatives.

To compute the derivatives of the $X_{ij}$, recall that an
isomonodromic deformation changes the matrix $A(z)$ in the
following way
\begin{eqnarray*}
{\rm d}A(z,T)=\p_zB(z,T)-\left[A(z,T),B(z,T)\right]
\end{eqnarray*}
The matrix 1-form $B(z,T)$ in this case is
\begin{eqnarray*}
B(z,T)=\sum_{i=1}^rB_{\al_i}(z,T){\rm
d}\al_i+\sum_{i=1}^nB_{c_i}(z,T){\rm d}c_i \\
\end{eqnarray*}
where the $B_{\al_i}(z,T)$ and the $B_{c_i}(z,T)$ are as follows
\cite{h94}
\begin{eqnarray}\label{eq:Bmatrix}
B_{\al_i}(z,T)&=&-{A_{\al_i}\over {(z-\al_i)}} \\
B_{c_i}(z,T)&=&zE_i+\sum_{j\neq
i}\sum_{k=1}^r{{E_iA_{\al_k}E_j+E_jA_{\al_k}E_i}\over
{\al_i-\al_j}}
\end{eqnarray}
where $E_i$ are the $n\times n$ diagonal matrices with 1 on the
$i^{th}$ entry and zero elsewhere.

The key observation here is that
$\p_zB_{\al_i}=\p_{\al_i}^eA(z,T)$ and
$\p_zB_{c_i}=\p_{c_i}^eA(z,T)$, where $\p_{\al_i}^e$ and
$\p_{c_i}^e$ denotes the explicit derivatives with respect to
$\al_i$ and $c_i$.

Therefore we have
\begin{eqnarray*}
\p_iA(z,T)=\p_i^eA(z,T)-\left[A(z,T),B_i(z,T)\right]
\end{eqnarray*}
where the index $i$ will be used to denote $\al_i$ or $c_i$.

Since the commutator term $\left[A(z,T),B_i(z,T)\right]$ fixes the
spectral invariance of the matrix $A(z,T)$ and it is known that
the coefficients $X_{ij}(T)$ of the spectral curve are spectral
invariance, \cite{h94} we see that the derivatives $\p_TX_{ij}$ in
(\ref{eq:zfix}) or (\ref{eq:wfix}) are the same as their explicit
derivatives with respect to the times $T$. That is
\begin{eqnarray*}
\p_TX_{ij}=\p_T^eX_{ij}
\end{eqnarray*}
this observation simplifies the calculations significantly as the
explicit dependence of the coefficients $X_{ij}$ on $\al_i$ and
$c_i$ are polynomials, which can be seen from (\ref{eq:spectral}).

In summary, the isomonodromic flows and the Whitham flows are
related by
\begin{proposition}\label{pro:isowhit}
Let $A(z)$ be a matrix of the form
\begin{eqnarray*}
A(z)=G^T(\Delta-z)^{-1}F+C
\end{eqnarray*}
where $G$, $F$ are $n\times r$ matrices constant in $z$,
$\Delta=\diag(\al_1,\ldots,\al_r)$ and $C=\diag(c_1,\ldots,c_n)$.

Define the spectral curve of $A(z)$ by
\begin{eqnarray*}
\det \left[ \mathbb{M}- \pmatrix{
  z & 0 \cr
  0 & w\cr} \right]=0
\end{eqnarray*}
where $\mathbb{M}$ is the following matrix
\begin{eqnarray*}
\mathbb{M}=\pmatrix{
  \Delta & -F \cr
  G^t & C\cr}
\end{eqnarray*}
then, when the system of linear ODE
\begin{eqnarray*}
{{dY}\over{dz}}=A(z,T)Y
\end{eqnarray*}
deforms isomonodromically with deformation parameters $\al_i$ and
$c_i$, the deformations of the spectral curve in
$\mathbb{CP}^1\times\mathbb{CP}^1$ can be expressed in terms of
the Whitham flows as follows
\begin{eqnarray}\label{eq:isowhit}
\p_{\al_i}&=&\p_{\al_i,1}+\sum_{k=1}^g\oint_{a_k}\left(\sum_{i=0}^r\sum_{j=0}^n\p_{\al_i}^eX_{ij}(T)z^iw^j\right)
P_w^{-1}dz\p_{h,k} \nonumber\\
\p_{c_i}&=&\p_{c_i,1}+\sum_{k=1}^g\oint_{a_k}\left(\sum_{i=0}^r\sum_{j=0}^n\p_{c_i}^eX_{ij}(T)z^iw^j\right)
P_w^{-1}dz\p_{h,k}
\end{eqnarray}
where $\p_i^e$ denotes the explicit derivative and $P_w$ is the
$w$ derivative of $P(z,w,T)$.

Here the twistor space of the Whitham hierarchy is taken to be the
set of algebraic curves given by the zero locus of polynomials
$P(z,w)$ in $\mathbb{CP}^1\times\mathbb{CP}^1$, where $P(z,w)$ has
degree at most $n$ in $w$ and $r$ in $z$. The marked points are
the points where $z$ or $w$ is infinity.
\end{proposition}

\paragraph*{Remark} As one could see from (\ref{eq:isowhit}) that,
the contraction $\p_i\hook\Pi$ of $\p_i$ with the 2-form $dz\wedge
dw$ would result in a 1-form on the curve that is holomorphic
everywhere apart from a second order pole at the point
$(z=\al_i,w=\infty)$ for $\p_{\al_i}$ or $(z=\infty,w=c_i)$ for
$\p_{c_i}$. In \cite{ta98a}, \cite{ta98b}, Takasaki rose the
question of whether it is possible to choose canonical basis of
cycles that could make these 1-forms normalized. This is the same
as choosing a basis of cycles so that the summation parts in
(\ref{eq:isowhit}) vanish. We do not know the answer but these
formula would simplify the problem significantly since all the
explicit derivatives can be computed.

We would now give an example.
\begin{example}{Painlev\'e V}\label{ex:PV}
\end{example}
The Painlev\'e V equation is the following second order nonlinear
ODE
\begin{eqnarray*}
u^{\prime\prime}=\left(u^{\prime}\right)^2\left({1\over{2u}}+{1\over{u-1}}\right)-{{u^{\prime}}\over{t}}
+{{\left(u-1\right)^2}\over{t^2}}\left(\al
u+{{\beta}\over{u}}\right)+{{\gamma u}\over{t}}+{{\delta
u(u+1)}\over{u-1}}
\end{eqnarray*}
where $\al$, $\beta$, $\gamma$ and $\delta$ are constants. It can
be represented as the isomonodromic deformations of a system of
linear ODE in which $A(z)$ is of the following form \cite{jmms}
\begin{eqnarray*}
A(z)={{A_1}\over{z-1}}+{{A_0}\over{z}}+\pmatrix{
  t & 0 \cr
  0& -t\cr}
\end{eqnarray*}
We will follow the approach in \cite{h94} to construct the
spectral curve. We first set the constant $\delta$ to be
$\delta=2$. In this case, the matrices $A_0$ and $A_1$ will be of
rank 1 and a factorization in (\ref{eq:factor}) can be carried
out.

In \cite{h94}, the Painlev\'e V was represented by the
isomonodromic deformations of the system of linear ODEs
\begin{eqnarray*}
{{dY}\over{dz}}Y^{-1}&=&\pmatrix{
  t & 0 \cr
  0& -t\cr} +2z^{-1}\pmatrix{
  -x_1y_1-\mu_1 & -y_1^2+{{\mu_1^2}\over{x_1^2}} \cr
  x_1^2 & x_1y_1-\mu_1\cr} \\&+&2(z-1)^{-1}\pmatrix{
  -x_2y_2-\mu_2 & -y_2^2+{{\mu_2^2}\over{x_2^2}} \cr
  x_2^2 & x_2y_2-\mu_2\cr}
\end{eqnarray*}
in which $\mu_1$, $\mu_2$ and $c={1\over 2}(x_1y_1+x_2y_2)$ are
constants. These constants are related to $\al$, $\beta$ and
$\gamma$ as follows
\begin{eqnarray*}
\al={{\mu_2^2}\over{2}},\quad\beta=-{{\mu_1^2}\over{2}},\quad\gamma=4c+2
\end{eqnarray*}
The solution $y$ to the Painlev\'e V equation is given by
$u=-{{x_1^2}\over{x_2^2}}$.

The factorization (\ref{eq:factor}) is given by
\begin{eqnarray*}
F={1\over{\sqrt{2}}}\pmatrix{
  x_1& y_1-{{\mu_1}\over{x_1}} \cr
  x_2& y_2-{{\mu_2}\over{x_2}}\cr} ,\quad G={1\over{\sqrt{2}}}\pmatrix{
  y_1+{{\mu_1}\over{x_1}}& -x_1 \cr
  y_2+{{\mu_2}\over{x_2}}& -x_2\cr}
\end{eqnarray*}
with $\Delta=\diag(0,1)$.

The spectral curve is then
\begin{eqnarray*}
P(z,w)&=&z^2w^2-w^2z+(\mu_1)wz-(\mu_1+\mu_2)w-t^2z^2+(t^2+tx_1y_1)z
\\ &-&2ct+{1\over
4}\left((x_1y_2-x_2y_1)^2+\left({{\mu_1^2}\over{u}}+\mu_2^2u\right)\right)=0
\end{eqnarray*}
This curve is of genus 1. The fixed points are $(z=0,w=\infty)$,
$(z=1,w=\infty)$, $(z=\infty,w=t)$ and $(z=\infty,w=-t)$. We can
now apply (\ref{eq:isowhit}) to express $\p_t$ in terms of the
Whitham flows
\begin{eqnarray*}
\p_t=\p_{t,1}+\oint_{a}\left(-2tz^2+(2t+x_1y_1)z-2c\right)
P_w^{-1}dz\p_{h,1}
\end{eqnarray*}
where $a$ is an $a$-cycle chosen on the curve.

\section*{Acknowledgements} Part of this work was carried out
under the supervision of N Woodhouse and was funded by the
Croucher Foundation. I would also like to thank L Mason for many
fruitful discussions.

\end{document}